\documentclass[12pt]{article}
 \usepackage{latexsym}
 \usepackage{amssymb}
\newcommand{\be}{\begin{equation}}
\newcommand{\ee}{\end{equation}}
\newcommand{\nn}{\nonumber}
\newcommand{\bea}{\begin{eqnarray}}
\newcommand{\eea}{\end{eqnarray}}

\newcommand{\la}{\langle}
\newcommand{\ra}{\rangle}

\usepackage{graphicx}%

\title{Beam-size or MD-effect at colliders and correlations
of particles in a beam}
\author{G.L.~Kotkin and V.G.~Serbo\thanks{E-mail: serbo@math.nsc.ru}\\
 {\it Novosibirsk State University, 630090, Novosibirsk, Russia}}
 \date{July 18, 2003}

\begin{document}
 \maketitle

\begin{abstract}

For several processes at colliding beams, macroscopically large
impact parameters give an essential contribution to the standard
cross section. These impact parameters may be much larger than the
transverse sizes of the colliding bunches. In that case, the
standard calculations have to be essentially modify. The
corresponding formulae for such a beam-size effect were given
twenty years ago without taking into account correlations of
particle coordinates in the beams. In the present paper we derive
formulae which necessary to take into account quantitatively the
effect of particle correlations in the spectrum of bremsstrahlung
as well as in pair production. Besides, we consider critically
recent papers~\cite{BK02,BK03} in which it was calculated a new
additional ``subtraction term'' related to the coherent
contribution into beam-size effect. We show that this result is
groundless and point out the origin of the mistake.

\end{abstract}



\section{\small INTRODUCTION}

The so called beam-size or MD-effect is a phenomenon discovered in
experiments \cite{Blinov82} at the MD-1 detector (the VEPP-4
accelerator with $e^+e^-$ colliding beams , Novosibirsk 1981). It
was found out that for ordinary bremsstrahlung, macroscopically
large impact parameters should be taken into consideration. These
impact parameters may be much larger than the transverse sizes of
the interacting particle bunches. In that case, the standard
calculations, which do not take into account this fact, will give
incorrect results. The detailed description of the MD-effect can
be found in review \cite{KSS}.

In $1980$--$1981$ a dedicated study of the process $e^+ e^-
\rightarrow e^+ e^- \gamma$ has been performed at the collider
VEPP-$4$ in Novosibirsk using the detector MD-1 for an energy of
the electron and positron beams $E_e=E_p = 1.8$ GeV and in a wide
interval of the photon energy $E_\gamma$ from $0.5$ MeV to
$E_\gamma \approx E_e$. It was observed \cite{Blinov82} that the
number of measured photons was smaller than that expected. The
deviation from the standard calculation reached $30 \%$ in the
region of small photon energies and vanished for large energies of
the photons. Yu.A.~Tikhonov \cite{Tikhonov82} pointed out that
those impact parameters $\varrho$, which give an essential
contribution to the standard cross section, reach values of
$\varrho_m \sim 5$ cm whereas the transverse size of the bunch is
$\sigma_\perp \sim 10^{-3}$ cm. The limitation of the impact
parameters to values $\varrho \lesssim \sigma_\perp$ is just the
reason for the decreasing number of observed photons.

The first calculations of this effect have been performed in Refs.
\cite{BKS} and \cite{BD} using different versions of
quasi--classical calculations in the region of large impact
parameters. Further experiments, including the measurement of the
radiation probability as function of the beam parameters,
supported the concept that the effect arises from the limitation
of the impact parameters. Later on, the effect of limited impact
parameters was taken into account when the single bremsstrahlung
was used for measuring the luminosity at the VEPP--$4$
collider~\cite{Blinov88} and at the LEP-I collider~\cite{Bini94}.
In the case of the VEPP--$4$ experiment~\cite{Blinov88}, it was
checked that the luminosities, obtained using either this process
or using other reactions (such as the double bremsstrahlung
process $e^+e^- \rightarrow e^+ e^- \gamma \gamma$, where the
MD-effect is absent), agreed with each other.

A general scheme  to calculate the finite beam size effect had
been developed in paper~\cite{KPS85a} starting from the quantum
description of collisions as an interaction of wave packets
forming bunches. Since the effect under discussion is dominated by
small momentum transfer, the general formulae can be considerably
simplified. The corresponding approximate formulae were given
in~\cite{KPS85a}. In a second step, the transverse motion of the
particles in the beams can be neglected. The less exact (but
simpler) formulae, which are then found, correspond to the results
of Refs.~\cite{BKS} and \cite{BD}. It has also been shown that
similar effects have to be expected for several other reactions
such as bremsstrahlung for colliding $ep$--beams~\cite{KPS85b},
\cite{KPSS88}, $e^+e^-$-- pair production in $e^\pm e$ and $\gamma
e$ collisions~\cite{KPS85a}. The corresponding corrections to the
standard formulae are now included in programs for simulation of
events at linear colliders. The influence of MD-effect on
polarization had been considered in Ref.~\cite{KKSS89}. In 1995
the MD-effect was experimentally  observed at the electron-proton
collider HERA~\cite{Piot95} on the level predicted
in~\cite{KPSS88}.

The possibility to create high-energy colliding $\mu^+\mu^-$ beams
is now wildly discussed. For several processes at such colliders a
new type of beam-size effect will take place --- the so called
linear beam-size effect~\cite{KMS}. The calculation of this effect
had been performed by method developed for MD-effect
in~\cite{KPS85a}.

It was realized in last years that  MD-effect in bremsstrahlung
plays important role for the problem of beam lifetime. At storage
rings TRISTAN and LEP-I, the process of a single bremsstrahlung
was the dominant mechanism for the particle losses in beams. If
electron loses more than $1\;\%$ of its energy, it leaves the
beam. Since  MD-effect reduced considerable the effective cross
section of this process, the calculated beam lifetime in these
storage rings was larger by about $25 \; \%$ for
TRISTAN~\cite{Funakoshi} and by about $40 \; \%$ for
LEP-I~\cite{Burkhardt} (in accordance with the experimental data)
then without taken into account the MD-effect. According to our
calculations~\cite{KS02}, at B-factories PEP-II and KEKB the
MD-effect reduces beam losses due to brems\-strah\-lung by about
20\%.

It is seen from this brief listing that the MD-effect is a
phenomenon interesting from the theoretical point of view and
important from the experimental point of view. In the present
paper we consider once again the MD-effect having in mind two
aims. The main purpose is to take into account correlations of
particle coordinates in the beams. Usually these correlations are
small, however, more accurate measurement may be sensitive to
them. In the present paper we derive formulae which necessary to
take into account quantitatively the effect of particle
correlations in the spectrum of bremsstrahlung as well as in pair
production.

Besides, we would like to consider critically recent
papers~\cite{BK02,BK03} in which previous
results~\cite{BKS,BD,KPS85a} about bremsstrahlung spectrum had
been revised. It was claimed that an additional subtraction
related to the coherent contribution has to be done. Analysis,
performed in paper~\cite{BK02}, results in the conclusion that
this additional ``subtraction term'' in the spectrum is not
important for the MD-1 experiment~\cite{Blinov82}, but it should
be taken into account in processing the HERA
experiment~\cite{Piot95}; it also may be important for the future
experiments at linear $e^+e^-$ colliders. It should be noted that
in paper~\cite{BK02} there is no derivation of the starting
formulae. On the other hand, in paper~\cite{BK02} there is a
general remark that their consideration was motivated by
corresponding calculations for bremsstrahlung of
ultra-relativistic electrons on oriented crystals. In our critical
remark~\cite{KS02} we had shown that the starting formulae
of~\cite{BK02} are incorrect. After that new paper~\cite{BK03} was
appeared in which there is the ``derivation'' of the starting
formulae used in~\cite{BK02}. Unfortunately, this derivation is
incorrect as well.

In the present paper we analyze the coherent and incoherent
contributions in the conditions, considered in
papers~\cite{BK02,BK03}, when the coherent length $l_{\rm coh}$ is
much smaller than the bunch length $l$ but much larger than the
mean distance between particles $a$, i.e. $a\ll l_{\rm coh} \ll
l$. We derive expressions for the coherent and incoherent
contributions and show that under these conditions the coherent
contribution is completely negligible and, therefore, there is no
need to revise the previous results. This conclusion is quite
natural. A usual bunch at colliders can be considered as a gaseous
media with a smooth particle distribution which has characteristic
scales of the order of bunch sizes. In particular, the average
particle density in such a bunch has the only scale in the
longitudinal direction --- the length of the bunch $l$. Therefore,
the average field of the bunch has the spectral components in the
region of frequencies $\omega=q_z c \sim c/l_{\rm coh} \lesssim
c/l$ and vanishes in the region of much higher frequencies
considered here. On the contrary, in the crystal case there is
another scale related to the size of the particle localization in
the crystal structure. In this case, the additional subtraction
should be taken into account for incoherent contribution. It seems
that the electron radiation on oriented crystals played a
misleading role for consideration of the MD-effect
in~\cite{BK02,BK03}. To clarify a question we give our
calculations in full details and  pointed out the origin of the
corresponding mistake in~\cite{BK02,BK03}.

In next Section we present the qualitative description of the
MD-effect. In Sect. III we discuss our approximations. Basic
formulae for coherent and incoherent contributions are given in
Sect. IV. Correction to the standard cross section, related to the
particle correlations, are derived in Sect. V. Our critical
remarks about results of papers~\cite{BK02,BK03} are presented in
Sect. VI. Some conclusions are given in Sect. VII.

\section{\small QUALITATIVE DESCRIPTION OF THE MD-EFFECT}

\begin{figure}[!t]
  \centering
  \setlength{\unitlength}{1cm}
\unitlength=2.0mm \special{em:linewidth 0.4pt}
\linethickness{0.4pt}
\begin{picture}(26.00,15.00)

\put(1.00,1.80){\line(1,0){24.00}}
\put(1.00,1.60){\line(1,0){24.00}}
\put(25.2,1.70){\vector(1,0){0.10}}

\put(1.00,9.20){\vector(1,0){6.00}}
\put(14.00,10.00){\circle{3.40}} \put(7.00,9.2){\line(1,0){5.3}}
\put(15.70,9.20){\vector(1,0){10.00}}
\put(15.0,11.5){\line(1,0){0.8}} \put(17,11.5){\line(1,0){0.8}}
\put(19,11.5){\line(1,0){0.8}} \put(21,11.5){\line(1,0){0.8}}
\put(23,11.5){\vector(1,0){2.00}}

\put(14.00,4.50){\vector(0,1){2.00}}
\put(14.00,2.90){\line(0,1){0.71}}
\put(14.00,1.80){\line(0,1){0.51}}
\put(14.00,7.50){\line(0,1){0.7}}

\put(1.00,10.4){\makebox(0,0)[cc]{$E_e$}}
\put(1.00,3.00){\makebox(0,0)[cc]{$E_p$}}
\put(22.00,12.70){\makebox(0,0)[cc]{$E_\gamma$}}
\put(22.00,5.00){\makebox(0,0)[cc]{$\hbar q=(\hbar\omega/c,\,
\hbar{\bf q})$}}

\end{picture}
    \caption{Block diagram of radiation by the electron.}
 \label{fig:1}
  \end{figure}
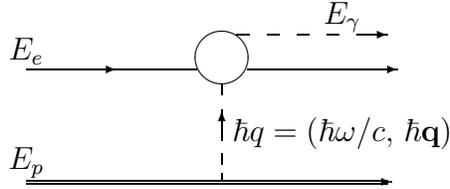

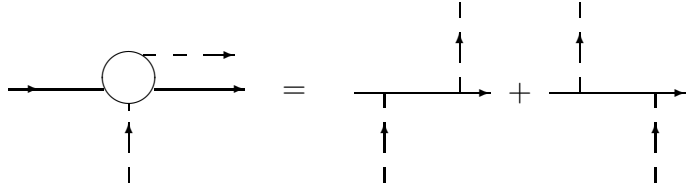
\begin{figure}[!t]
  \centering
\unitlength=2.00mm \special{em:linewidth 0.4pt}
\linethickness{0.4pt}
\begin{picture}(47.00,15.00)
\put(2.00,9.20){\vector(1,0){2.00}}
\put(10.00,10.00){\circle{3.40}} \put(3.00,9.20){\line(1,0){5.30}}
\put(11.70,9.20){\vector(1,0){6.00}}
\put(10.00,5.00){\vector(0,1){2.00}}
\put(10.00,3.00){\line(0,1){1.00}}
\put(10.00,7.80){\line(0,1){0.50}}
\
\
\put(11.00,11.50){\line(1,0){0.80}}
\put(13.00,11.50){\line(1,0){0.80}}
\put(15.00,11.50){\vector(1,0){2.00}}
\
\put(25.00,9.00){\vector(1,0){9.00}}
\put(38.00,9.00){\vector(1,0){9.00}}

\put(27.00,8.00){\line(0,1){1.00}}
\put(27.00,5.00){\vector(0,1){2.00}}
\put(27.00,3.00){\line(0,1){1.00}}

\put(32.00,9.00){\line(0,1){1.00}}
\put(32.00,11.00){\vector(0,1){2.00}}
\put(32.00,14.00){\line(0,1){1.00}}

\put(40.00,9.00){\line(0,1){1.00}}
\put(40.00,11.00){\vector(0,1){2.00}}
\put(40.00,14.00){\line(0,1){1.00}}

\put(45.00,8.00){\line(0,1){1.00}}
\put(45.00,5.00){\vector(0,1){2.00}}
\put(45.00,3.00){\line(0,1){1.00}}

\put(21.00,9.00){\makebox(0,0)[cc]{=}}
\put(36.00,9.00){\makebox(0,0)[cc]{+}}

\end{picture}
 \caption{Compton scattering of equivalent photon on the electron.}
 \label{fig:2}
\end{figure}

Qualitatively we describe the MD--effect using as an example the
$e p \rightarrow e p \gamma$ process\footnote{Below we use the
following notations: $N_e$ and $N_p$ are the numbers of electrons
and protons (positrons) in the bunches, $\sigma_z=l$ is the
longitudinal, $\sigma_x$ and $\sigma_y$ are the horizontal and
vertical transverse sizes of the proton (positron) bunch,
$\gamma_e=E_e/(m_ec^2)$, $\gamma_p=E_p/(m_pc^2)$ and $r_e=e^2/(m_e
c^2)$ is the classical electron radius.}. This reaction is defined
by the diagrams of Fig.~\ref{fig:1} which describe the radiation
of the photon by the electron (the contribution of the photon
radiation by the proton can be neglected). The large impact
parameters $\varrho \gtrsim \sigma_\perp$, where $\sigma_\perp$ is
the transverse beam size, correspond to small momentum transfer
$\hbar q_\perp \sim (\hbar / \varrho) \lesssim (\hbar /
\sigma_\perp)$. In this region, the given reaction can be
represented as a Compton scattering (Fig.~\ref{fig:2}) of the
equivalent photon, radiated by the proton, on the electron. The
equivalent photons with frequency $\omega$ form a ``disk'' of
radius $\varrho_m \sim \gamma_p c / \omega$ where $\gamma_p = E_p
/ (m_p c^2)$ is the Lorentz-factor of the proton. Indeed, the
electromagnetic field of the proton is $\gamma_p$--times
contracted in the direction of motion. Therefore, at distance
$\varrho$ from the axis of motion a characteristic longitudinal
length of a region occupied by the field can be estimated as
$\lambda \sim \varrho / \gamma_p$ which leads to the frequency
$\omega \sim c / \lambda \sim \gamma_p c / \varrho$.

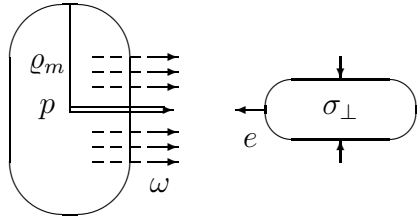
\begin{figure}[!b]
  \centering
\unitlength=2.00mm \special{em:linewidth 0.4pt}
\linethickness{0.4pt}
\begin{picture}(47.00,15.00)

\put(13.00,7.00){\oval( 8,14)}

\put(13.00 ,6.85){\line(1,0){6.30}}
 \put(13.00,7.15){\line(1,0){6.30}}
 \put(13.00 ,6.90){\line(0,1){7.00}}
\put(19.40 ,7.00){\vector(1,0){0.50}}

\put(14.50 ,8.50){\line(1,0){0.70}}
 \put(15.70,8.50){\line(1,0){0.70}}
  \put(17.00,8.50){\line(1,0){0.70}}
 \put(18.20 ,8.50){\vector(1,0){2.00}}

\put(14.50 ,9.50){\line(1,0){0.70}}
\put(15.70,9.50){\line(1,0){0.70}}
 \put(17.00,9.50){\line(1,0){0.70}}
\put(18.20 ,9.50){\vector(1,0){2.00}}

\put(14.50 ,10.50){\line(1,0){0.70}}
 \put(15.70,10.50){\line(1,0){0.70}}
  \put(17.00,10.50){\line(1,0){0.70}}
\put(18.20 ,10.50){\vector(1,0){2.00}}

\put(14.50 ,5.50){\line(1,0){0.70}}
 \put(15.70,5.50){\line(1,0){0.70}}
  \put(17.00,5.50){\line(1,0){0.70}}
  \put(18.20 ,5.50){\vector(1,0){2.00}}

\put(14.50 ,4.50){\line(1,0){0.70}}
 \put(15.70,4.50){\line(1,0){0.70}}
  \put(17.00,4.50){\line(1,0){0.70}}
 \put(18.20 ,4.50){\vector(1,0){2.00}}

\put(14.50 ,3.50){\line(1,0){0.70}}
 \put(15.70,3.50){\line(1,0){0.70}}
  \put(17.00,3.50){\line(1,0){0.70}}
\put(18.20 ,3.50){\vector(1,0){2.00}}

\put(11.50,10.00){\makebox(0,0)[cc]{$\varrho_m$}}
\put(11.50,7.00){\makebox(0,0)[cc]{$p$}}
\put(19.00,2.00){\makebox(0,0)[cc]{$\omega$}}

\put(31.00,7.00){\oval( 10,4)}

\put(26.00,7.00){\vector(-1,0){2.00}}
\put(31.00,10.50){\vector(0,-1){1.50}}
\put(31.00,3.50){\vector(0,1){1.50}}

\put(31.00,7.00){\makebox(0,0)[cc]{$\sigma_\perp$}}
\put(25.00,5.00){\makebox(0,0)[cc]{$e$}}

\end{picture}
\caption{Scattering of equivalent photons, forming the ``disk"
with radius $\varrho_m=\gamma_p c/\omega$, on the electron beam
with radius $\sigma_\perp$. }
 \label{fig:3}
\end{figure}

In the reference frame connected with the collider, the equivalent
photon with energy $\hbar \omega$ and the electron with energy
$E_e \gg \hbar \omega$ move toward each other (Fig.~\ref{fig:3})
and perform a Compton scattering. The characteristics of this
process are well known. The main contribution to the Compton
scattering is given by the region where the scattered photons fly
in a direction opposite to that of the initial photons. For such a
backward scattering, the energy of the equivalent photon $\hbar
\omega$ and the energy of the final photon $E_\gamma$ and its
emission angle $\theta_\gamma$ are related by
 \begin{equation}
    \hbar \omega = \frac{E_\gamma}{4 \gamma^2_e (1 - E_\gamma/E_e )}
   \left[1+ (\gamma_e\theta_\gamma)^2 \right]
 \label{1.1a}
 \end{equation}
and, therefore, for typical emission angles $\theta_\gamma\lesssim
1/\gamma_e$ one has
\begin{equation}
 \hbar \omega \sim {E_\gamma \over 4 \gamma^2_e (1 - E_\gamma/E_e
   )}\,.
 \label{1.1}
 \end{equation}

As a result, we find the radius of the ``disk'' of equivalent
photons with the frequency $\omega$ (corresponding to a final
photon with energy $E_\gamma$) as follows:
 \begin{equation}
 \varrho_m = {\gamma_p c\over  \omega} \sim \lambda_e\; {4 \gamma_e
\gamma_p}\, {E_e - E_\gamma \over E_\gamma} \,,\;\;\;
\lambda_e={\hbar\over m_ec}=3.86\cdot 10^{-11}\; {\rm cm}\,.
 \label{size}
 \end{equation}
For the HERA collider with $E_p=820$ GeV and $E_e=28$ GeV one
obtains
 \begin{equation}
 \varrho_m \gtrsim 1\; {\rm cm \ \ \ for\ \ \ } E_\gamma \lesssim
0.2 \;{\rm GeV}\ .
 \label{1.3}
 \end{equation}
Equation (\ref{size}) is also valid for the $e^- e^+ \rightarrow
e^- e^+ \gamma$ process with replacement protons by positrons. For
the VEPP-4 collider it leads to
 \begin{equation}
 \varrho_m \gtrsim 1\; {\rm cm \ \ \ for\ \ \ } E_\gamma \lesssim
15\; \mbox{ MeV }\,,
  \label{1.4}
  \end{equation}
for the PEP-II and KEKB colliders we have
 \begin{equation}
 \varrho_m \gtrsim 1\; {\rm cm \ \ \ for\ \ \ } E_\gamma \lesssim
0.1\; \mbox{ GeV }\,.
  \label{1.4a}
  \end{equation}

The  standard  calculation corresponds to the interaction of the
photons forming the ``disk'' with the unbounded flux of electrons.
However, the particle beams at the HERA collider have finite
transverse beam sizes of the order of $\sigma_\perp\sim 10^{-2}$
cm. Therefore, the equivalent photons from the region
$\sigma_\perp \lesssim \varrho \lesssim \varrho_m$ cannot interact
with the electrons from the other beam. This leads to the
decreasing number of the observed photons and the  ``observed
cross section''  $d \sigma_{\rm obs}$ is smaller than the
standard  cross section $d \sigma$ calculated for an infinite
transverse extension of the electron beam,
 \begin{equation}
   d \sigma_{\rm obs} = d \sigma - d \sigma_{\rm cor}.
  \label{1.5}
  \end{equation}
Here the correction $d \sigma_{\rm cor}$ can be presented in the
form
 \begin{equation}
 d \sigma_{\rm cor} = d \sigma_{\rm C}(\omega,\,E_e,\,E_\gamma) \
dn(\omega)
  \label{1.6}
  \end{equation}
where $dn(\omega)$ denotes the number of ``missing'' equivalent
photons and $d \sigma_{\rm C}$ is the cross section of the Compton
scattering.  Let us stress that the equivalent photon
approximation in this region has a high accuracy (the neglected
terms are of the order of $1/\gamma_p$). But for the qualitative
description it is sufficient to use the logarithmic approximation
in which this number is (see~\cite{BLP}, \S 99)
 \begin{equation}
  dn = {\alpha \over \pi} {d\omega \over \omega} {d q_{\perp}^2 \over
q_{\perp}^2} \,.
 \label{1.7}
 \end{equation}
Since $q_\perp  \sim 1 / \varrho$, we can present the number of
``missing'' equivalent photons in the form
 \begin{equation}
 dn = {\alpha \over \pi} {d \omega \over \omega} { d\varrho^2 \over
\varrho^2}
 \label{1.8}
 \end{equation}
with the integration region in $\varrho$:
 \begin{equation}
 \sigma_\perp \lesssim \varrho \lesssim \varrho_m = {\gamma_p c
\over \omega}\,.
 \label{1.9}
 \end{equation}
As a result, this number is equal to
 \begin{equation}
 dn(\omega) = 2 {\alpha \over \pi} { d\omega \over \omega} \ln
{\varrho_m \over \sigma_\perp }\,,
 \label{1.10}
 \end{equation}
and the correction to the standard cross section with logarithmic
accuracy is\footnote{Within this approximation, the standard cross
section has the form
 $$
d\sigma = d \sigma_{\rm C} {\alpha \over \pi} {d\omega \over
\omega} {d q_{\perp}^2 \over q_{\perp}^2}={16\over 3} \alpha
r^2_e\, {dy\over y}\, \left(1-y+\mbox{${3\over 4}$} y^2\right)
\ln{4\gamma_e \gamma_p (1-y)\over y }
 $$
with the integration region $\omega /(c \gamma_p) \lesssim q_\perp
\lesssim m_e c/\hbar$ corresponding to the impact parameters
$\varrho$ in the interval $\lambda_e \lesssim \varrho \lesssim
\varrho_{m}$. }
 \begin{equation}
  d\sigma_{\rm cor} = {16\over 3} \alpha r^2_e\, {dy\over y}\,
\left(1-y+\mbox{${3\over 4}$} y^2\right) \ln{4\gamma_e \gamma_p
(1-y)\lambda_e\over y \sigma_\perp}\,, \;\;y={E_\gamma\over
E_e}\,.
  \label{1.11}
 \end{equation}

\section{\small APPROXIMATIONS}

For future linear  $e^+ e^-$ colliders the transverse sizes of the
beams will change significantly during the time of interaction due
to a mutual attraction of very dense beams. However, for most of
the ordinary accelerators, including practically all $e^+e^-$ and
$ep$ storage rings, the change of the transverse beam sizes during
the collisions can be neglected. Below we use two main
approximations: 1) the particle movement in the bunches has a
quasi-classical character; 2) the particle distribution remains
practically unchanged during the collision. For definiteness, we
use again the $ep$ collisions as an example.

Therefore, if the proton (electron) bunch moves along (opposite)
the direction of $z$-axis with the velocity $v_p$ ($v_e$), its
density has the form
 \begin{equation}
 n_p=n_p ({\mbox{\boldmath $\varrho$}}, z-v_pt)\,,\;\;
n_e=n_e ({\mbox{\boldmath $\varrho$}}, z+v_et)\,.
 \label{15}
 \end{equation}
We also introduce so called ``transverse densities''
 \begin{equation}
 n_p ({\mbox{\boldmath $\varrho$}})= \int n_p\,dz\,,\;\;
n_e ({\mbox{\boldmath $\varrho$}})= \int n_e\,dz
 \label{16}
 \end{equation}
which is equal to the total number of protons (electrons) which
cross a unit area around the impact parameter ${\mbox{\boldmath
$\varrho$}}$ during the collision. Using the transverse densities,
we express the luminosity for collisions of beams whose axes
separated by impact parameter ${\mbox{\boldmath $\varrho$}}$ as
 \be
L({\mbox{\boldmath $\varrho$}})= \int n_e({\bf r}_\perp)\,n_p
({\mbox{\boldmath $\varrho$}}+{\bf r}_\perp)\, d^2 {\bf
r}_\perp\,.
 \label{lumr}
 \ee
The usual luminosity for a single collision of $ep$-beams $L_{ep}$
is then
 \be
 L_{ep} = L(0)\,.
 \label{lumu}
 \ee

Below we consider in detail the case when an electron deflection
angle $\theta_e$ is smaller than the typical radiation angle $\sim
1/\gamma_e$. It is easy to estimate the ratio of these angles. The
electric {\bf E} and magnetic {\bf B} fields of the proton bunch
are approximately equal in magnitude, $|{\bf E}| \approx  |{\bf
B}| \sim eN_p /(\sigma_x+\sigma_y)$. These fields are transverse
and they deflect the electron into the same direction. In such
fields the electron moves around a circumference of  radius $R\sim
\gamma_e m_e c^2/(eB)$ and gets the deflection angle $\theta_e
\sim l/R$. Therefore, the ratio of these angles is of the order of
\begin{equation}
{ \theta_e \over (1/\gamma_e)} \sim \eta={r_e N_p\over
\sigma_x+\sigma_y}  \,.
 \label{1}
\end{equation}
The parameter $\eta \gg 1$ only for the SLC and future linear
$e^+e^-$ colliders, in most of the colliders $\eta \lesssim 1$.

In our consideration we use the equivalent photon approximation.
In the region of interest (where impact parameters are large,
$\varrho \gtrsim \sigma_\perp$) this simple and transparent method
has a high accuracy: the neglected terms are of the order of
$1/\gamma$. It should be stressed that the operator
quasi-classical method, used in Ref.~\cite{BK02}, just coincides
in this region with the equivalent photon approximation.

\section{\small COHERENT AND INCOHERENT CONTRIBUTIONS}

\subsection{General formulae}

The corresponding formulae for the number of events in a single
collision of the electron and proton bunches can be found in
papers~\cite{Ginzyaf}, \cite{ESS1}. To calculate the MD-effect, we
need to know the distribution of equivalent photons (EP) for large
values of impact parameters. In this region we can consider the
electron--proton scattering as the scattering of electrons on the
electromagnetic field of the proton bunch. Replacing this field by
the flux of EP with some frequency distribution, we obtain the
number of events in the form
\begin{equation}
dN = dL_{\gamma e }(\omega ) \  d\sigma_{\rm C} (\omega,E_e,\,
E_\gamma )\,, \;\;\; dL_{\gamma e}(\omega )= n_\gamma
({\mbox{\boldmath $\varrho$}}, \omega) d\omega \,n_e
({\mbox{\boldmath $\varrho$}}) \, d^2\varrho \,.
 \label{18}
\end{equation}
Here $n_e ({\mbox{\boldmath $\varrho$}})$ is the transverse
electron density (\ref{16}) and $n_\gamma ({\mbox{\boldmath
$\varrho$}},\omega) d\omega$ is  the transverse density of EP with
the frequencies in the interval from $\omega$ to $\omega + d\omega
$. The quantity $dL_{\gamma e}(\omega )$ denotes the differential
luminosity for the collisions of EP and electrons and $d
\sigma_{\rm C} (\omega,E_e,\, E_\gamma)$ is the Compton cross
section for the scattering of the equivalent photon with the
frequency $\omega$ on the electron.

For comparison with the experimental data the number of events in
a single collision of beams $dN$ should be averaged over many
collisions of bunches in a given experiment. For example, the
typical rate at the HERA collider is less then $1/100$
bremsstrahlung photons in a certain interval of frequencies per a
single collision of the beams, therefore, in that experiment the
averaging over many collisions of bunches really does exist.

The transverse density of the EP is determined by density of the
electromagnetic field for a given frequency, i.e. by  $\mid {\bf
E}_\omega ({\mbox{\boldmath $\varrho$}})\mid ^2/(4\pi)$, where
${\bf E}_\omega ({\mbox{\boldmath $\varrho$}})$ is the spectral
component of the collective electric field of the proton bunch. As
a result, the transverse density of the EP is
 \begin{equation}
n_\gamma ({\mbox{\boldmath$\varrho$}},
 \omega )\;d\omega ={c\over 4\pi^2}\;\left\langle \mid {\bf E}_\omega
({\mbox{\boldmath $\varrho$}})\mid ^2\right\rangle\;{d\omega \over
\hbar \omega}\,,
 \label{18a}
 \end{equation}
where the sign $\langle \dots \rangle$ denotes the above mentioned
statistical averaging. The field ${\bf E}_\omega ({\mbox{\boldmath
$\varrho$}})$ itself depends on a distribution of charges in the
proton bunch at $t=0$. We introduce the exact (fluctuating)
density of the proton bunch $n({\bf r})$ and the averaging density
 \begin{equation}
  n_p({\bf r})= \langle n({\bf r}) \rangle
 \end{equation}
as well as the corresponding form factor
\begin{equation}
 F_p({\bf q})=  \int
n_p({\bf r})\;\mbox {e}^{-{\rm i}{\bf qr}}\; d^3r
\end{equation}
with the normalization
\begin{equation}
F_p(0)=\int n_p({\bf r})\, d^3r =N_p \ .
\end{equation}
In the classical limit
 \begin{equation}
  n({\bf r})=\sum_{a}\,\delta({\bf r}- {\bf r}_a)\,,
 \label{22a}
 \end{equation}
where ${\bf r}_a$ is the radius-vector of the $a$-th proton. In
these notations, the exact (fluctuating) collective field is
 \begin{equation}
 {\bf E}_\omega ({\mbox{\boldmath $\varrho$}})=- { {\rm i} e \over
\pi c}\, \int  d^2 q_\bot\,{{\bf q}_\bot \mbox {e}^{{\rm i}{\bf
q}_\bot {\mbox{\scriptsize\boldmath $\varrho$}}} \over D({\bf q})}
\, \int d^3r\,n({\bf r})\,\mbox{e}^{-{\rm i}{\bf qr}}\,,\;\;
D({\bf q})={\bf q}^2_\bot+{q_z^2\over \gamma_p^2}
 \label{12b}
\end{equation}
with $q_z= \omega/c$.

As a result, the number of events
 \begin{equation}
 dN  \propto n_\gamma ({\mbox{\boldmath$\varrho$}},
 \omega ) ={\alpha\over 4\pi^4 \omega}
\int\, {({\bf q}_\bot {\bf q}'_\bot)\over D({\bf q})\,D({\bf q}')}
\; \mbox{e}^{{\rm i}({\bf q}_\bot -{\bf q}'_\bot)
\mbox{\scriptsize\boldmath $\varrho$}}\, S({\bf q},{\bf q}') \;
d^2 q_\bot\,d^2 q'_\bot\;
 \end{equation}
depends on the beam structure factor
 \begin{equation}
 S({\bf q},{\bf q}') =
\int S({\bf r},{\bf r}')\;\mbox {e}^{-{\rm i}({\bf qr} -{\bf
q}'{\bf r}')}\,d^3r\, d^3r'\,, \;\;
 S({\bf r},{\bf r}')=\left\langle \,n({\bf r})\, n({\bf r}')\,\right\rangle
 \label{sum}
 \end{equation}
in which
 \begin{equation}
  q_z=q'_z= \omega/c\,.
 \label{qo}
 \end{equation}

Below we analyze these formulae in conditions when the coherence
length $l_{\rm coh} \sim c/\omega$ is much smaller than the bunch
length $l$, but much larger that the mean distance between
particles in the beam $a$, i. e. at
 \be
 a\ll {c\over \omega} \sim  l_{\rm coh} = {4 \gamma_e^2 \hbar c\over
 E_\gamma}\,(1-E_\gamma/E_e) \ll l.
 \label{27a}
 \ee

\subsection{The beam structure factor}

The obtained general formulae include the  coherent and incoherent
contributions. The coherent contribution is determined by the
average field which is given  by Eq. (\ref{12b}) with the
replacement of the exact density $n({\bf r})$ by the average
density $n_p({\bf r})$. The averaged density of the proton bunch
has a single scale in the longitudinal direction --- the length of
the bunch $l$. Therefore, the average field of the bunch is
essential in the region of frequencies $\omega =cq_z\lesssim c/l$
and should be small in the region of large frequencies $\omega \gg
c/l$. In particular, if the proton bunch has the Gaussian
distribution,
 \be
n_p({\bf r})= {N_p\over (2\pi)^{3/2} \sigma_x \sigma_y l}\,
\exp\left[ -{x^2\over 2\sigma_x^2}- {y^2\over 2\sigma_y^2}-
{z^2\over 2l^2}\right]\,,
 \ee
its form factor is equal to
\begin{equation}
F_p({\bf q})=N_p\; \mbox{exp} \left[ - \mbox{${1\over
2}$}(q_x\sigma_{x})^2 - \mbox{${1\over 2}$} (q_y\sigma_{y})^2 -
 \mbox{${1\over 2}$}
(\omega l/c)^2 \right]
 \label{G}
\end{equation}
and vanishes in the discussed region of frequencies from the
interval (\ref{27a}).

If we introduce the density fluctuation
 \be
 \Delta n({\bf r})=n({\bf r})-n_p({\bf r})\,,
 \ee
we can rewrite the average product of densities in the form
 \be
\left\langle n({\bf r})\,n({\bf r}')\right\rangle= n_p({\bf
r})\,n_p({\bf r}') +\left\langle\Delta n({\bf r})\,\Delta n({\bf
r}')\right\rangle\,.
 \ee
In accordance with this presentation, we split the function $
S({\bf r},{\bf r}')$ in two items called coherent and incoherent
contribution:
 \be
S=S_{\rm coh}+S_{\rm incoh}\,,\;\;
 S_{\rm coh}({\bf r},{\bf r}')=n_p({\bf r})n_p({\bf r}')\,,\;\;
S_{\rm incoh}({\bf r},{\bf r}')=\left\langle\Delta n({\bf
r})\,\Delta n({\bf r}')\right\rangle\,.
 \ee
The coherent contribution to the structure factor is equal to
 \begin{equation}
  S_{\rm coh}({\bf q},{\bf q}')= F_p({\bf q})\,F_p^*({\bf q}')\,.
 \label{COH}
  \end{equation}
This formula was used in Refs.~\cite{Ginzyaf}, \cite{ESS1} to
obtain main characteristics of the coherent bremsstrahlung. It
also allows us to obtain the following estimate for the Gaussian
beam in the region of interest (at $|q_x|,\, |q'_x|\lesssim 1/
\sigma_{x}$ and $|q_y|,\,|q'_y| \lesssim 1/ \sigma_{y}$):
 \begin{equation}
 S_{\rm coh}({\bf q},{\bf q}')\sim N_p^2\,\mbox{exp} \left[ - (\omega l/c)^2
\right]\,.
 \label{cc}
 \end{equation}

Let us now consider the incoherent contribution. A bunch at
colliders can be treated as a continues media with a smooth
average particle distribution. It was shown in Appendix that for
such a media the function $ S_{\rm incoh}({\bf r},{\bf r}')$ is
expressed only via the average density ${n}_p({\bf r})$ and via
the correlation function $C({\bf r},{\bf r}')$ as follows
\begin{equation}
 S_{\rm incoh}({\bf r},{\bf r}')=\delta({\bf r}- {\bf r}')\,{n}_p({\bf
 r})+C({\bf r},{\bf r}')\,.
 \label{sigma}
 \end{equation}

If we neglect the correlations of the particle coordinates in the
beam, the correlation function $C({\bf r},{\bf r}')$ vanishes, and
we obtain (taking into account Eq. (\ref{qo}))
 \begin{equation}
 S_{\rm incoh}({\bf q},{\bf q}')= F({\bf q}_\bot -{\bf q}'_\bot)\,.
 \label{INC}
 \end{equation}
It is important that this expression is determined only by the
transverse average density of the proton bunch and it does not
depend on $\omega$. Formula (\ref{INC}) has been used to derive
the previous results about MD-effect (for details see
review~\cite{KSS} and Sect. V). For the Gaussian beam in the
region of interest, we get from (\ref{INC}) a useful estimate
 \begin{equation}
 S_{\rm incoh}({\bf q},{\bf q}')  \sim N_p\,.
 \end{equation}

The correlations of the particle coordinates may arise due to
Coulomb interaction of particles in the beam. In this case the
characteristic quantity --- the correlation length $l_{\rm
correl}$ --- is related to the Debay radius. It is evident that
the correlations are negligible if the correlation length is much
larger then the coherence length, i.e. at $l_{\rm correl} \gg
l_{\rm coh}$. According to an estimate \cite{BD} it is just the
case for the VEPP-4 experiment \cite{Blinov82}. In any case, the
correlations may give an essential correction to the standard
bremsstrahlung cross section only if
 \be
l_{\rm correl} \lesssim l_{\rm coh}\,.
  \ee
Therefore, the important quantity is the spectral component of the
correlation function:
 \be
C_\omega({\bf r}_\perp, \, {\bf r}'_\perp)= \int C({\bf r}, \,
{\bf r}')\, {\rm e}^{-{\rm i} \omega(z-z')/c}\, dz\,dz'\,.
 \label{coro}
 \ee
With this notation the incoherent contribution is now:
 \begin{equation}
 S_{\rm incoh}({\bf q},{\bf q}')= F({\bf q}_\perp-{\bf q}'_\perp)+
C_\omega({\bf q}_\perp, \, {\bf q}'_\perp)\,,
 \label{INCa}
 \ee
where
 \be
C_\omega({\bf q}_\perp, \, {\bf q}'_\perp)= \int C_\omega({\bf
r}_\perp, \, {\bf r}'_\perp) \, \mbox{e}^{-{\rm i}({\bf q}_\bot
{\bf r}_\perp-{\bf q}'_\bot {\bf r}'_\perp)}\,d^2 r_\bot\,d^2
r'_\bot\,.
 \label{INC2}
 \end{equation}

\section{\small CORRECTION TO THE STANDARD BREMSSTRAHLUNG CROSS SECTION}

Let us compare the coherent and incoherent contributions for the
Gaussian beams. In this case, the ratio
 \begin{equation}
 {dN^{\rm coh}\over dN^{\rm incoh}} \sim {S_{\rm coh}({\bf q},{\bf
 q}')\over S_{\rm incoh}({\bf q},{\bf q}')} \sim
 N_p\,\mbox{exp} \left[ - (\omega l/c)^2
\right]
 \label{comprison}
 \end{equation}
is determined by the parameters $\omega l/c$. Since $ \hbar \omega
\sim E_\gamma /[4 \gamma_e^2 (1-E_\gamma/E_e)]$, it is also useful
to introduced the coherence length (\ref{27a}) and the critical
energy for the coherent bremsstrahlung
 \begin{equation}
  E_c ={4 \gamma_e^2 \hbar c\over l}\,.
 \end{equation}
If the coherence length is large, $l_{\rm coh}\gtrsim l$, or if
the final photon energy is small, $E_\gamma \lesssim E_c$, the
parameter $\omega l/c \lesssim 1$ and the coherent contribution is
dominant.

On the contrary, in the region of large photon energy, $E_\gamma
\gg E_c$, or small coherence length, $l_{\rm coh}\ll l$,
considered here, the incoherent contribution dominates. In
particular,  for $N_p \sim 10^{11}$ the ratio $dN^{\rm
coh}/dN^{\rm incoh}$ is small even for $\omega l/c = 6$,
 \begin{equation}
 {dN^{\rm coh}\over dN^{\rm incoh}} \sim
 N_p\,\mbox{e}^{ - 36} \ll 1\,,
 \label{comprison2}
 \end{equation}
and the coherent contribution becomes completely negligible. In
this case the number of events for bremsstrahlung can be presented
in the form (cf. (\ref{1.5}))
 \be
dN^{\rm incoh}= L_{ep}\, d \sigma_{\rm obs}\,,\;\;
 d \sigma_{\rm obs} = d \sigma - d \sigma_{\rm cor}\,,
 \label{correc}
 \ee
where $L_{ep}$ is the luminosity (\ref{lumu}) of the
$ep$-collisions, $d \sigma$ is the standard cross section for the
$ep\to ep\gamma$ process and $d \sigma_{\rm cor}$ is the
correction related to the MD-effect. Then we perform integration
over ${\bf q}_\perp$ and ${\bf q}'_\perp$ using the well known
equality
 \begin{equation}
 \int  {{\bf q}_\perp \, \mbox{e}^{{\rm i}{\bf q}_\bot
\mbox{\scriptsize\boldmath $\varrho$}} \over {\bf q}^2_\perp
+(1/b)^{2}}\, d^2q_\perp = {2\pi {\rm i}\over b}\,{\mbox{\boldmath
$\varrho$} \over \varrho} \,K_1(\varrho/b)
 \label{K1}
 \end{equation}
where $K_n(x)$ denotes the modified Bessel function of third kind
with integer index $n$ (McDonald function). As a resul, we obtain
the correction to the standard cross section in the form (cf. with
the approximate formulae (\ref{1.6}), (\ref{1.10}))
 \begin{equation}
d \sigma_{\rm cor}= d\sigma_C(\omega, E_e, E_\gamma)\, {\alpha
\over \pi} { d\omega \over \omega}\,G(\omega)\,,
 \label{xcorr}
 \ee
where $d\sigma_C$ is the Compton cross section and the function
$G(\omega)$ consists of two items
 \be
G(\omega)=G^{(1)}(\omega)+G^{(2)}(\omega)\,.
 \ee
The first item represents the previous result for the MD-effect
(without taking into account correlations),
 \be
G^{(1)}(\omega)= \int {d^2\varrho \over \pi \varrho_m^2} \left[1-
{L({\mbox{\boldmath $\varrho$}})\over L(0)} \right]\,
K^2_1(\varrho/\varrho_m)\,,\;\; \varrho_m = {c\gamma_p\over
\omega}\,,
 \label{prev}
 \ee
where $L({\mbox{\boldmath $\varrho$}})$ is defined in
(\ref{lumr}). Some other useful expressions for $G^{(1)}(\omega)$
as well as its asymptotics can be found in \cite{KSS}. The second
item is directly related to the correlation function
(\ref{sigma}), (\ref{coro}):
 \be
G^{(2)}(\omega)= -\int {d^2\varrho \over \pi \varrho_m^2}\,
{n_e({\mbox{\boldmath $\varrho$}})\over L_{ep}}\, C_\omega({\bf
r}_\perp +{\mbox{\boldmath $\varrho$}},{\bf r}'_\perp
+{\mbox{\boldmath $\varrho$}})\, {({\bf r}_\perp {\bf
r}'_\perp)\over r_\perp r'_\perp}\, K_1(r_\perp/\varrho_m)
K_1(r'_\perp/\varrho_m) \,d^2r_\perp d^2r'_\perp\,.
 \label{new}
 \ee
Note that the main contribution to $G^{(1)}(\omega)$ (\ref{prev})
is given by the region of large impact parameters (\ref{1.9})
while the main contribution to $G^{(2)}(\omega)$ is given by the
region $\varrho\sim \sigma_\perp$.

The quantity $d\sigma_C \,d\omega/\omega$ in (\ref{xcorr}) can be
expressed via the energy $E_\gamma$ and the emission angle
$\theta_\gamma$ of the final photon as follows (taking into
account relation (\ref{1.1a}))
 \be
d\sigma_C\, {\alpha \over \pi} { d\omega \over \omega}= 2\alpha\,
r^2_e \,{dy\over y} {dz\over (1+z)^2} \,F(y,z)\,,\;\; F(y,z)=
2(1-y){1+z^2\over (1+z)^2}+y^2\,,
 \ee
where
 \be
y={E_\gamma \over E_e}\,, \;\; z= \left(\theta_\gamma
\gamma_e\right)^2\,,\;\; r_e = {e^2\over m_ec^2}\,.
 \ee

\section{\small CRITICAL REMARKS ABOUT RESULTS of Refs.~\cite{BK02,BK03}}

We derive the final expression for the incoherent contribution
from general equations (\ref{18}), (\ref{18a}) and (\ref{12b}) as
a simple consequence of natural assumptions about the particle
distribution in a proton bunch. It is useful to rewrite these
equations in the form convenient for comparison with the
corresponding equations in~\cite{BK02,BK03}. To do this, we note
that the Compton cross section $d\sigma_{\rm C}\propto |{\bf
e}{\bf M}_{\rm Compton} |^2$ where ${\bf e}{\bf M}_{\rm Compton}$
is the amplitude of the Compton scattering for the EP with the
polarization vector ${\bf e}={\bf E}_\omega ({\mbox{\boldmath
$\varrho$}})/\mid {\bf E}_\omega ({\mbox{\boldmath
$\varrho$}})\mid$. Therefore, the number of events in a given
collisions of beams is proportional to $|M|^2$, where
\begin{equation}
M={\bf E}_\omega ({\mbox{\boldmath $\varrho$}}){\bf M}_{\rm
Compton}
 \end{equation}
is related to the probability amplitude of the process. Further,
we use Eqs. (\ref{22a}) and (\ref{K1}) and present the collective
field of the proton bunch ${\bf E}_\omega ({\mbox{\boldmath
$\varrho$}})$ as a sum of fields of all protons:
 \be
{\bf E}_\omega ({\mbox{\boldmath $\varrho$}})=\sum_{a=1}^{N_p}
{\bf E}^{(a)}_\omega ({\mbox{\boldmath $\varrho$}})\,,\;\; {\bf
E}^{(a)}_\omega ({\mbox{\boldmath $\varrho$}})={2e\over c
\varrho_m}\, {\mbox{\boldmath $\varrho$}'_a\over \varrho'_a}
K_1\left(\varrho'_a / \varrho_m\right)\, {\rm e}^{-{\rm i} \omega
z_a/c}\,,
 \label{Ea}
 \end{equation}
where $\mbox{\boldmath $\varrho$}'_a=\mbox{\boldmath
$\varrho$}-\mbox{\boldmath $\varrho$}_a$ is the impact parameter
between the electron and the $a$-th proton and the parameter
$\varrho_m=\gamma_pc/\omega$ is the radius of the ``disc'' of EP
(see Fig.~3).

As a consequence, the amplitude $M$ is the sum
 \begin{equation}
 M=\sum_{a=1}^{N_p} m_a\, {\rm e}^{-{\rm i} \omega
z_a/c}\,,\;\;\; m_a= {2e\over c \varrho_m}\, K_1\left(\varrho'_a /
\varrho_m\right) \, {\mbox{\boldmath $\varrho$}'_a\over
\varrho'_a}{\bf M}_{\rm Compton}\,,
 \label{Ma}
 \end{equation}
where the item $m_a \exp(-{\rm i} \omega z_a/c)$ is the
contribution to $M$ related to the interaction of the electron
with the $a$-th proton, while $|M|^2$ can be presented as the
double sum
 \be
|M|^2= \sum_{a,b} m_am_b^*\, {\rm e}^{-{\rm i} \omega
(z_a-z_b)/c}\,.
 \label{ds}
 \ee
We split this sum into sum with $a=b$ and sum with $a\neq b$:
 \be
|M|^2= \Sigma_1+ \Sigma_2\,,\;\; \Sigma_1= \sum_a |m_a|^2\,,\;\;
\Sigma_2=\sum_{a\neq b} m_am_b^*\, {\rm e}^{-{\rm i} \omega
(z_a-z_b)/c}\,.
 \label{split}
 \ee
Equations (\ref{Ma})---(\ref{split}) can be considered as the same
starting formulae in our approach, based on the equivalent photon
approximation, and in approach of~\cite{BK02,BK03}, based on the
operator quasi-classical method. However, further calculations are
quite different. For simplicity, below we consider the case when
we can neglect the correlations between the particle coordinates
in the proton bunch.

In our approach, the number of events is proportional to $|M|^2$
averaged over collisions of beams, i.e.
 \be
dN\propto \langle |M|^2 \rangle = \langle \Sigma_1 \rangle +
\langle \Sigma_2 \rangle
 \ee
In the considered region of large frequencies (\ref{27a}), the
item $\langle \Sigma_2 \rangle$, corresponding to the coherent
contribution, vanishes,
 \be
|\langle \Sigma_2 \rangle|\ll \langle \Sigma_1 \rangle\,.
 \label{s2s1}
 \ee
Since $m_a$ does not depend on the longitudinal coordinate $z_a$,
the average value of $|m_a|^2$ is determined by the transverse
average density of the proton bunch (\ref{16}),
 \be
\langle|m_a|^2 \rangle= \int |m_a|^2\,{n_p({\bf r}_a)\over N_p}\,
d^3r_a=\int |m_a|^2\,{n_p(\mbox{\boldmath $\varrho$}_a)\over
N_p}\, d^2\varrho_a\,,
 \ee
and it does not depend on index $a$. Therefore, the item
 \be
\langle \Sigma_1 \rangle= N_p\,\langle|m_a|^2 \rangle
 \label{S1ma}
 \ee
leads to the correction, corresponding to $G^{(1)}$ in
(\ref{xcorr})---(\ref{prev}).

Authors of~\cite{BK02,BK03} as the first step had performed
averaging over transverse coordinates of the protons.  Certainly,
after that they get the same expression for $\Sigma_1$ as in
(\ref{S1ma}). For $\Sigma_2$ they had obtained the following
expression
 \be
\langle \Sigma_2 \rangle_\perp =|\langle m_a \rangle_\perp|^2\,
Z\,,\;\; Z=\sum_{a\neq b} {\rm e}^{-{\rm i} \omega (z_a-z_b)/c}\,,
 \label{Z}
 \ee
where
 \be
\langle m_a \rangle_\perp =\int m_a\,{n_p(\mbox{\boldmath
$\varrho$}_a)\over N_p}\, d^2\varrho_a\,.
 \ee
The principal mistake in~\cite{BK03} consists in the incorrect
calculation of $Z$. It is not difficult to understand its true
behavior in the considered region  of large frequencies
(\ref{27a}). The quantity $Z$ fluctuates near zero for various
sets of coordinates $\{z_a\}=z_1, z_2, ..., z_{N_p}$,
corresponding to various collisions of beams, and after averaging
over many collisions one obtains the estimate (\ref{s2s1}). This
natural behavior of $Z$ is confirmed by numerical calculations
given below.

When calculating $Z$, the authors of~\cite{BK03} add and subtract
the items with $a= b$, as a consequence,
 \be
Z= J -N_p \,,\;\; J=\sum_{a, b} {\rm e}^{-{\rm i} \omega
(z_a-z_b)/c}=\left|\sum_{a} {\rm e}^{-{\rm i} \omega
z_a/c}\right|^2\,.
 \label{Z21}
 \ee
Their next step consists in replacement the sum $J$ by the
integral
 \be
J\to \left|\,\int {\rm e}^{-{\rm i} \omega z/c}\, n_p({\bf r})\,
d^3r\,\right|^2\,,
 \label{mist}
  \ee
which is negligible in the considered region. In particular, for
the Gaussian beam the replacement (\ref{mist}) means the
following:
 \be
J\to N^2_p\,\mbox{exp} \left[ - (\omega l/c)^2 \right] \ll N_p
 \label{mist22}
 \ee
(cf. (\ref{cc}), (\ref{comprison}) and (\ref{comprison2})). This
estimate leads to a large negative value of
 \be
Z=-N_p
 \label{mist2}
  \ee
and to
 \be
\langle \Sigma_2 \rangle_\perp =-N_p\,|\langle m_a
\rangle_\perp|^2\,.
 \label{Z2}
 \ee
Just expression (\ref{Z2}) is a new ``subtraction term'' derived
in~\cite{BK02,BK03}.

The mistake of~\cite{BK03} consists in replacement (\ref{mist}).
This replacement is true for the region of small frequencies
$\omega l/c \ll 1$ when $J=N_p^2$ and $Z=-N_p+J=N_p(N_p-1)\approx
N_p^2$, but such a replacing is completely wrong in the considered
region of large frequencies (\ref{27a}). To show this, we perform
numerical calculation of the sum $J$. For a given collision of
beams we can consider a set of the longitudinal proton coordinates
$\{z_a\}$ as a set of random quantities with some distribution
$w(z)$. We assume below that
 \be
w(z) = {1\over \sqrt{2\pi}\,l}\, \exp\left(-{z^2\over
2l^2}\right)\,.
 \ee
Now the sum
 \be
\sum_{a=1}^{N_p}{\rm e}^{-{\rm i}qz_a} =C-{\rm i} S
 \ee
with
 \be
C=\sum_{a=1}^{N_p} \cos({qz_a})\,,\;\; S=\sum_{a=1}^{N_p}
\sin({qz_a})\,,\;\; q=\omega/c
 \label{73}
 \ee
is also the random quantity as well as
 \be
 J= C^2 +S^2\,.
 \label{74}
 \ee
The quantities $C$ and $S$ are the sums of large numbers of random
items. Therefore, one can expect that they distribute in
accordance with the normal law:
 \be
{dW\over dC}= {1\over \sqrt{2\pi\, N_p}\, \Delta c}
\,\exp\left[-{\left(C-N_p{\bar c}\right)^2\over 2N_p (\Delta c)^2}
\right]\,,\;\; {dW\over dS}= {1 \over \sqrt{2\pi\,
N_p{\overline{s^2}}}} \, \exp\left[-{S^2\over
2N_p{\overline{s^2}}} \right]\,,
 \label{75}
 \ee
where $\Delta c = \sqrt{\overline{c^2}-\overline{c}^2}$ and
 \bea
  \label{76}
{\bar c}&=&{\overline{\cos{(qz)}}}=\int_{-\infty}^{+\infty}
w(z)\cos(qz)\, dz={\rm e}^{-(ql)^2/2}\,,\;\;
 {\bar s}={\overline{\sin{(qz)}}}=0\,,
 \\
{\overline{c^2}}&=&{\overline{\cos^2{(qz)}}}= \mbox{${1\over
2}$}\left(1+{\rm e}^{-2(ql)^2} \right)\,,\;\;
 {\overline{s^2}}= {\overline{\sin^2{(qz)}}}=\mbox{${1\over 2}$}\left(1-{\rm
e}^{-2(ql)^2} \right)\,.
 \nn
 \eea
Numerical calculations had been performed using the generator of
random numbers from MATLAB. These calculations confirm the above
distributions (\ref{75}). In particular, it can be seen from
Fig.~4
and  Fig.~5
where the results of numerical calculations for $ql=1,\; N_p=10^2$
and $ql=10,\;N_p=10^3$ are presented for $10^4$ various sets of
$\{z_a\}$.

Since
 \be
 \la C \ra = N_p{\bar c}\,,\;\;
 \left\la \left(C-N_p{\bar c}\right)^2\right\ra
  = N_p({\overline{c^2}}-{\overline{c}}^2)\,,\;\;
  \la S \ra=0\,,\;\;
 \left\la S^2 \right\ra = N_p{\overline{s^2}}\,,
 \ee
we have
 \be
\la J \ra = N_p +N_p\left(N_p-1\right)\,\bar{c}^2\,,\;\; \la Z \ra
= N_p\left(N_p-1\right)\,\bar{c}^2 \approx N^2_p\,\mbox{exp}
\left[ - (\omega l/c)^2 \right]\,.
  \ee
Moreover, taking into account that in the considered case
$(N_p\,\bar{c})^2\ll N_p$, we find that (in contrast to
(\ref{mist22}), (\ref{mist2}))
 \be
\la J \ra = N_p\,,\;\;|\la Z \ra| \ll N_p\,.
 \ee
The distribution of the random quantity $J$ becomes very simple at
$ql\gg 1$:
 \be
 {dW\over dJ}= {1\over N_p}\, {\rm e}^{-J/N_p}\,.
 \label{distr}
 \ee
The results of numerical calculations, presented on Fig.~6
, confirm Eq. (\ref{distr}). It should be noted that distribution
(\ref{distr}) is rather wide,
 \be
 \Delta J =\sqrt{\la J^2 \ra -\la J \ra^2 }= N_p\,,
 \ee
therefore, the averaging over many various  sets of $\{z_a\}$ is
necessary to obtain the stable result for $\la J \ra$.

This consideration shows that the effect, derived
in~\cite{BK02,BK03}, is absent just in the region, discussed in
these papers. At the end of this section we reconsider the
experiments analyzed in paper~\cite{BK02}.

The HERA experiment~\cite{Piot95}. In this case $E_e= 27.5$ GeV
and $l= 8.5$ cm, therefore, $E_c = 27$ keV. For the observed
photon energies $E_\gamma =2\div 8$ GeV, the parameter
 \begin{equation}
 {\omega l\over c} \sim {E_\gamma\over E_c} > 10^{4}\,,
 \end{equation}
and the coherent contribution is completely negligible. Therefore,
the new correction to the previous results on the level of $10$
\%, obtained in~\cite{BK02}, is, in fact, absent.

The VEPP-4 experiment~\cite{Blinov82}. In this case $E_e= 1.84$
GeV and $l= 3$ cm, therefore, $E_c = 0.34$ keV. For the observed
photon energies $E_\gamma \gtrsim 1$ MeV, the parameter
 \begin{equation}
 {\omega l\over c} \sim {E_\gamma\over E_c} > 10^{3}\,,
 \end{equation}
and the coherent contribution is completely negligible.

The case of a ``typical linear collider'' with $E_e=500$ GeV and
$E_\gamma = E_e/1000$. This example,  considered in
paper~\cite{BK02}, is irrelevant for the discussed problem, since
the coherent radiation (called in this case {\it beamstrahlung})
at a typical linear collider absolutely dominates in this very
region over the ordinary incoherent bremsstrahlung --- see, for
example, the TESLA project~\cite{TDR} and Sect. 3 of~\cite{BK03}.

\section{\small CONCLUDING REMARKS}

In the present paper we had performed analysis of the coherent and
incoherent contributions to the bremsstrahlung in conditions
(\ref{27a}) when the incoherent contribution dominates but large
impact parameters give an essential contribution to the standard
cross section. In this conditions the known correction
(\ref{prev}) to the standard cross section is determined by the
transverse distribution of particles in the beams.

We take into account correlations of particles in the beam. The
corresponding correction to the standard cross section is given by
Eq. (\ref{new}) and it is determined by correlations of particles
in the transverse as well as in longitudinal coordinates.

Through the paper we consider MD-effect in bremsstrahlung. The
MD-effect for the $e^+e^-$ pair production (for example, in the
reaction $\gamma e\to e^+e^-e$) can be considered in the same
manner --- for detail see Sect 7.1 from review~\cite{KSS}.

We had shown that replacing the sum $J$ (\ref{Z21}) by the
integral (\ref{mist}) in conditions (\ref{27a}) is incorrect. As a
consequence, papers~\cite{BK02,BK03} are incorrect as well.

\section*{Acknowledgments}

We are very grateful to  A.~Chernykh, V.~Dmitriev, V.~Fadin,
I.~Ginzburg, V.~Katkov, I.~Kolokolov, A.~Milshtein and
N.~Vinokurov for useful discussions. This work is supported in
part by INTAS (code 00-00679), RFBR (code 03-02-17734) and by the
Fund of Russian Scientific Schools (code 2339.2003.2).

\section*{\small APPENDIX. AVERAGE PRODUCT OF PARTICLE DENSITIES IN THE
BEAM}

In the classical limit (\ref{22a}) the average product of particle
densities in the proton beam is given by the double sum over
protons in the beam
 \be
S({\bf r},{\bf r}')=\sum_{a,b=1}^{N_p}\langle \delta({\bf r}- {\bf
r}_a)\,\delta({\bf r}'- {\bf r}_b)\rangle\,.
 \ee
We split this expression into sum with $a=b$ and sum with $a\neq
b$:
 \be
S({\bf r},{\bf r}') =S_1+S_2\,,\;\;
 S_1=\delta({\bf r}- {\bf r}')\, \sum_a \langle \delta({\bf r}-
{\bf r}_a)\rangle\,,\;\; S_2=\sum_{a\neq b} \langle \delta({\bf
r}- {\bf r}_a)\,\delta({\bf r}'- {\bf r}_b)\rangle\,.
 \ee
To perform the averaging, we introduce the average proton
distribution function
 \be
f({\bf r})= n_p ({\bf r})/N_p
 \ee
with the normalization
 \be
\int f({\bf r})\,d^3 r =1\,.
 \ee
It gives
 \be
\langle \delta({\bf r}- {\bf r}_a)\rangle= \int \delta({\bf r}-
{\bf r}_a)\,f({\bf r}_a)\,d^3r_a=f({\bf r})\,.
 \ee
Note, that quantity $\langle \delta({\bf r}- {\bf r}_a)\rangle$
does not depend on index $a$ and, therefore,
 \be
S_1=\delta({\bf r}- {\bf r}')\,N_p\,f({\bf r})= \delta({\bf r}-
{\bf r}')\,n_p({\bf r})\,.
 \ee

If we can neglect correlations between the particle coordinates,
then the average product $\langle \delta({\bf r}- {\bf
r}_a)\,\delta({\bf r}'- {\bf r}_b)\rangle$  for $a\neq b$ can be
presented as the product of two averaged factors:
 \be
\langle \delta({\bf r}- {\bf r}_a)\,\delta({\bf r}'- {\bf
r}_b)\rangle=\langle \delta({\bf r}- {\bf
r}_a)\rangle\,\langle\delta({\bf r}'- {\bf r}_b)\rangle \;\;
\mbox{ for } \; a\neq b\,.
 \ee
As a consequence,
 \be
S_2=\sum_{a\neq b} \langle \delta({\bf r}- {\bf
r}_a)\rangle\,\langle\delta({\bf r}'- {\bf
r}_b)\rangle=N_p(N_p-1)\,f({\bf r})\,f({\bf r}')\,.
 \label{corre}
 \ee

If we do not neglect the correlations between the particle
positions, we should introduce the correlation function $C({\bf
r},{\bf r}')$ as follows
 \be
S_2=\sum_{a\neq b} \langle \delta({\bf r}- {\bf
r}_a)\rangle\,\langle\delta({\bf r}'- {\bf r}_b)\rangle + C({\bf
r},{\bf r}')\,.
 \ee
In that case we obtain instead of (\ref{corre}) the expression
 \be
S_2=N_p(N_p-1)\,f({\bf r})\,f({\bf r}')+ C({\bf r},{\bf r}') \,.
 \ee
As a result,
 \be
S({\bf r},{\bf r}')=N_p(N_p-1)\,f({\bf r})\,f({\bf
r}')+\delta({\bf r}- {\bf r}')\,n_p({\bf r})+ C({\bf r},{\bf r}')
\,.
 \ee
Since in right-hand-side of this equation the first and the second
items usually do not compensate each other, we can use
approximation
 \be
 N_p(N_p-1)\approx N_p^2
 \ee
and, therefore,
 \be
S({\bf r},{\bf r}')=n_p({\bf r})\,n_p({\bf r}')+\delta({\bf r}-
{\bf r}')\,n_p({\bf r})+ C({\bf r},{\bf r}') \,.
 \ee



\begin{figure}[h]
\begin {center}
\includegraphics[width=8.5cm]{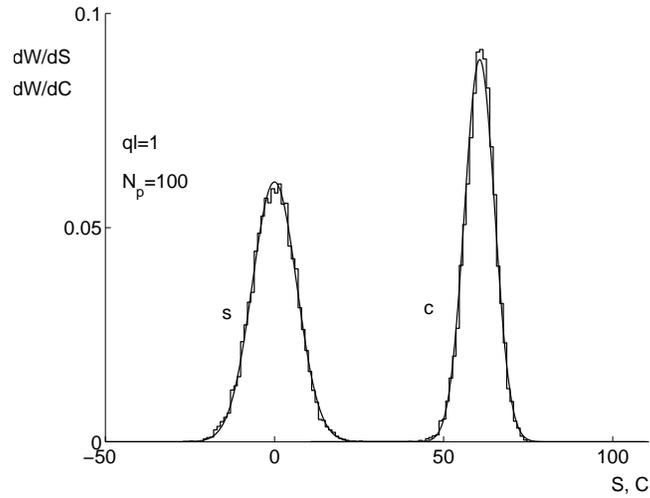}
\end{center}
\caption{The  curves $s$ and $c$ are given for $dW/dS$ and
$dW/dC$, respectively, in accordance with Eqs. (\ref{75}) and
(\ref{76}) for $ql=1$, $N_p=10^2$. The histograms represent
results of numerical calculations for distribution of $S$ and $C$
defined in (\ref{73}) for $10^4$ various sets of random numbers
$\{z_a/l\}$} \label{wsc}
\end{figure}


\begin{figure}[h]
\begin {center}
\includegraphics[width=8.5cm]{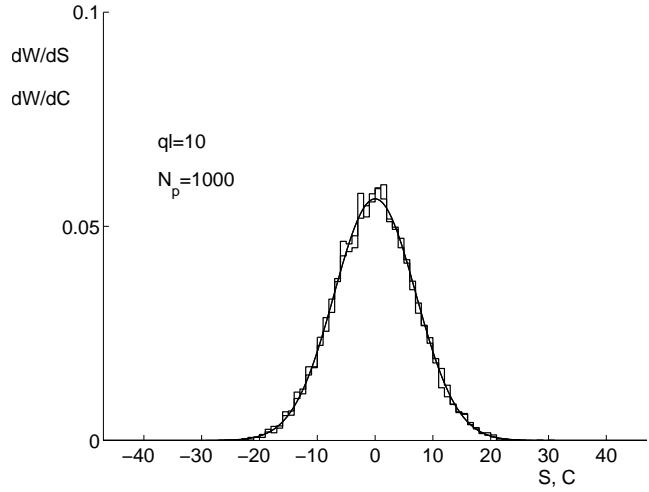}
\end{center}
\caption{The same as in Fig. 4, but for $ql=10$, $N_p=10^3$ (in
this case the curves $s$ and $c$ are practically coincide).}
  \label{wcs}
\end{figure}


\begin{figure}[h]
\begin{center}
\includegraphics[width=9cm]{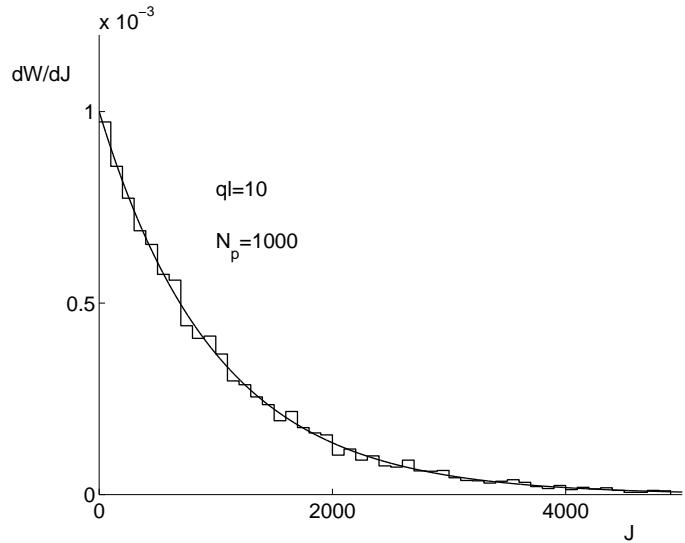}
\end{center}
\caption{The curve is given for $dW/dJ$ in accordance with Eq.
(\ref{distr}) for $ql=10$, $N_p=10^3$. The histogram represents
the result of numerical calculations for distribution of $J$
defined in (\ref{Z21}), (\ref{74}) for $10^4$ various sets of
random numbers $\{z_a/l\}$.}
 \label{dwdj}
\end{figure}

\end{document}